# Optimization of Spin Wave Propagation with Enhanced Group Velocities by Exchange Coupled Ferrimagnet/Ferromagnet Bilayers

K. An[1], V. S. Bhat[1], M. Mruczkiewicz[2], C. Dubs[3], and D. Grundler[1,4†]

[1]*Laboratory of Nanoscale Magnetic Materials and Magnonics, Institute of Materials (IMX), School of Engineering, École Polytechnique Fédérale de Lausanne (EPFL), 1015 Lausanne, Switzerland*
[2]*Institute of Electrical Engineering, Slovak Academy of Sciences, 841 04 Bratislava, Slovakia*
[3]*INNOVENT e.V. Technologieentwicklung, Prüssingstraße 27B, D-07745 Jena, Germany*
[4]*Institute of Microengineering (IMT), School of Engineering, École Polytechnique Fédérale de Lausanne (EPFL), 1015 Lausanne, Switzerland*

We report broadband spectroscopy and numerical analysis by which we explore propagating spin waves in a magnetic bilayer consisting of a 23 nm thick permalloy film deposited on 130 nm thick $Y_3Fe_5O_{12}$. In the bilayer, we observe a characteristic mode that exhibits a considerably larger group velocity at small in-plane magnetic field than both the magnetostatic and perpendicular standing spin waves. Using the finite element method, we confirm the observations by simulating the mode profiles and dispersion relations. They illustrate the hybridization of spin wave modes due to exchange coupling at the interface. The high-speed propagating mode found in the bilayer can be utilized to configure multi-frequency spin wave channels enhancing the performance of spin wave based logic devices.

I. Introduction

In the field of spintronics, building blocks for low-power consuming logic and data processing devices might exploit spin waves [1–3]. Studies on the generation, manipulation, and detection of spin waves have already evolved into a broader research field called magnonics [4–8]. One of the immediate challenges is the generation of short-waved spin waves, which are particularly important for miniaturizing devices [9,10]. Here, long propagation distance and high group velocity of spin waves are desired. In a magnetic thin film, perpendicular standing spin waves (PSSWs) reflect exchange-dominated modes which are quantized between the top and bottom surface. For a magnetic film with a thickness of 130 nm, the first order PSSW (PSSW1) corresponds to a large wave vector of $k_{perp} = 24.2$ rad/μm pointing perpendicular to the film. When PSSWs acquire a small in-plane wave vector component $k_i$ in e.g. ferrimagnetic yttrium iron garnet $Y_3Fe_5O_{12}$ (YIG), the modes are known to exhibit a small group velocity, contradicting fast and efficient signal transmission. Conventional micro-structured microwave antennas allow one to transfer $k_i$ on the order of 1 rad/μm but, if integrated to a planar film, are known to excite PSSW only weakly [11]. Recently, *undulation* of a CoFeB film was shown to allow for efficient PSSW excitation within the same film [12]. For *planar* YIG, it was reported that top layers of either Co [13] or $Co_{40}Fe_{40}B_{20}$ [14] enhanced PSSW amplitudes considerably. This was attributed to interfacial exchange-mediated spin transfer torque. Also, bilayers composed of permalloy (Py) and YIG have been studied [15–17] concerning magnetic resonances and coupling at the interface. However the propagation characteristics such as signal transmission and group velocities have not been analyzed yet.

In this letter, we explore propagating spin waves excited in bilayers of Py and YIG. In addition to the magnetostatic surface spin wave (MSSW), we detect an unexpectedly high group velocity ($v_g$) of about 4500 m/s in a small in-plane magnetic field $\mu_0 H$ of 4 mT. Finite-element modeling of the YIG/Py bilayer attributes the high $v_g$ to a PSSW-like hybridized mode with a total wave vector $k_{tot} = \sqrt{k_{perp}^2 + k_i^2} = 24.2$ rad/μm. Originating from a PSSW the hybrid mode resides at a different frequency compared to the MSSW with $k_{tot} = k_i$. Our observation allows one to configure multi-frequency magnonic devices offering spin wave modes with high group velocities.

II. Experimental and numerical method

A single-crystalline YIG film with a thickness of 130 nm was grown by a liquid phase epitaxy on a (111) gadolinium gallium garnet substrate [18]. A 23 nm thick Py ($Ni_{81}Fe_{19}$) film was deposited on the YIG using electron beam evaporation. As a control sample, a nominally identical YIG film without Py was explored. Subsequently, a 34 nm thick $SiO_2$ layer was deposited to provide electrical insulation and allow for the integration of two parallel coplanar wave guides (CPWs). They consisted of 180 nm-thick gold on a 5 nm thick adhesion layer of Ti. Both, the YIG and YIG/Py sample were shaped to parallelograms using ion-beam etching. This shape avoided the formation of standing spin waves in lateral direction. The signal and ground lines of CPWs were 3.3 μm wide and exhibited an edge-to-edge separation of 2.7 μm. The distance *d* between two signal lines amounted to 30 and 18 μm for YIG and YIG/Py sample, respectively. The in-plane wave vector most efficiently excited by the CPWs was determined to be $k_i = 0.5$ rad/μm via Fourier transforming the in-plane

[†]email: dirk.grundler@epfl.ch

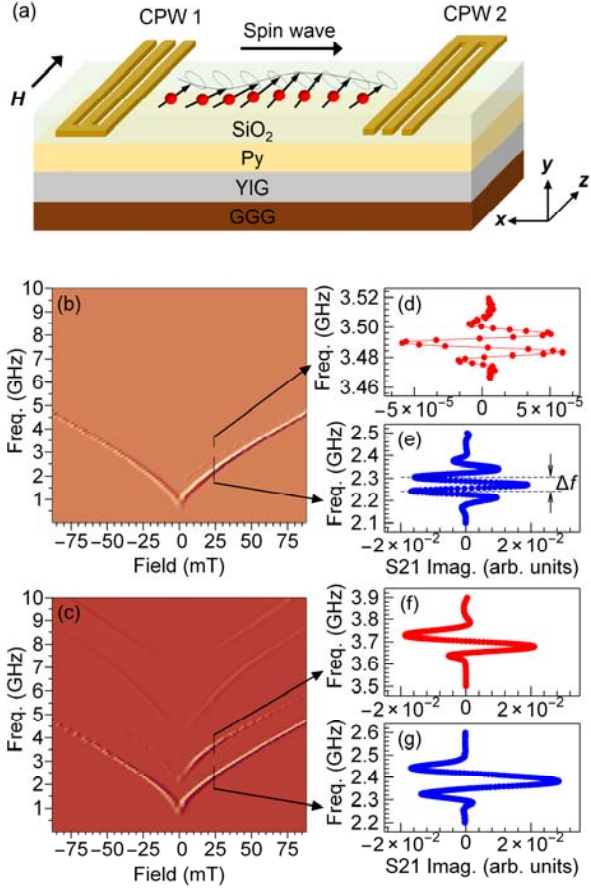

FIG. 1 (a) Schematics of sample and transmission measurement geometry. Measured spin wave transmission signal from (b) bare YIG film (c) YIG/Py bilayer. The arrows indicate where the linecuts of spectra are taken. The linecuts (d) and (e) represent the PSSW1 and MSSW in bare YIG at 24 mT. Similarly, the linecuts (f) and (g) represent the PSSW1 and MSSW in the YIG/Py bilayer at 24 mT. The separation between adjacent maximum frequencies are defined as $\Delta f$. An example of such frequency spacing is shown in (e).

component of the microwave field [19]. A microwave power of -5 dBm was provided by the vector network analyzer (VNA), and transmission signals from CPW1 to CPW2 (S21) were recorded. A schematic of the measurement geometry is shown in Fig. 1a. In our experiment, the spin wave propagates along the $-x$ direction, the $y$ axis is perpendicular to the film, and the magnetic field was applied along the $z$ axis. To enhance the signal-to-noise ratio spectra measured at two slightly different magnetic fields were subtracted from each other. The subtracted signal $\Delta S21$ is displayed in color-coded graphs as shown in Fig. 1b and 1c. Considering the phase sensitive detection of the VNA, the signal S21 at the receiver port experiences a phase accumulation of $k_i d$ due to spin waves propagating between CPW1 and CPW2. Spin waves with different $k_i$ arrive at the receiver port with different phases, resulting in an oscillating amplitude (as shown in the Fig. 1d-g). A phase difference of $2\pi$ is accumulated when the difference in wave vectors satisfies the relation $2\pi = d\Delta k_i$. Approximating the spin wave dispersion relation to be linear in $k_i$, we write $v_g = 2\pi df/dk \approx 2\pi \Delta f/\Delta k$. The group velocity is obtained by $v_g = d\Delta f$ [20,21].

Using the finite element method COMSOL, we solved the Landau–Lifshitz equation (without damping term). For YIG, we used the exchange constant $A_{YIG} = 3.7 \times 10^{-12}$ J/m [22], the gyromagnetic ratio $\gamma_{YIG}/(2\pi) = 28$ GHz/T and the magnetization value of $\mu_0 M_{YIG} = 0.1835$ T extracted from a separately conducted magnetic resonance measurement. For Py, we used $A_{Py} = 1.03 \times 10^{-11}$ J/m [23], $\gamma_{Py}/(2\pi) = 29.4$ GHz/T [24], and $\mu_0 M_{Py} = 1.018$ T [24]. We imposed a periodic boundary condition along the $x$ axis with a unit cell size of 20 nm. Considering the periodic boundary condition, the dynamic components of magnetization, $m_x$ and $m_y$, take the form of a plane wave, i.e., $m_j(x, y, t) = m_j(y) e^{-i(kx-\omega t)}$ with $j = x, y$. The wave vector $k$ is parametrically swept within the simulation to obtain the dispersion relation. Two 1 nm thick intermediate layers were introduced at the interface to simulate the interfacial exchange coupling between YIG and Py. We used the same magnetic parameters corresponding to their parent layers, with an interfacial exchange constant defined by $A_{int} \equiv (A_{YIG} + A_{Py})/2$.

III. Results

Transmission spectra (imaginary part of $\Delta S21$) obtained from the bare YIG film and the YIG/Py bilayer are shown in Fig. 1b and 1c, respectively. Compared to the bare YIG (Fig. 1b), more spin wave branches are visible in the YIG/Py bilayer (Fig. 1c). The lowest and most prominent branch in each graph is attributed to the MSSW in YIG. The signal strength is larger for positive than for negative field, attributed to nonreciprocity [25]. In Fig. 1c a second branch is pronounced residing at larger frequency compared to the MSSW. The higher frequency is consistent with the PSSW1 in YIG as will be discussed later. Above about 6 GHz, we observe two further modes for the YIG/Py bilayer. Their frequencies are close to both the MSSW in Py and second order PSSW in YIG. An anti- crossing at about 53 mT in Fig.

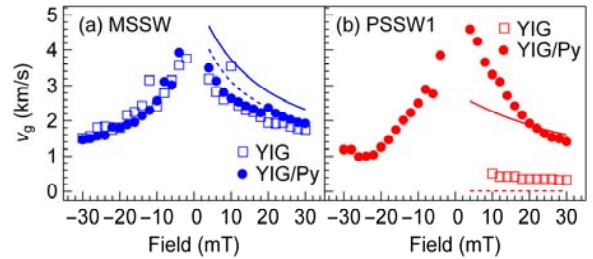

FIG. 2 Field dependencies of group velocities for (a) MSSW and (b) PSSW1, both attributed to YIG. The solid (dashed) lines are simulation results for the YIG/Py (bare YIG) sample. The PSSW1 group velocity was not extracted for negative magnetic field due to low signal to noise ratio.

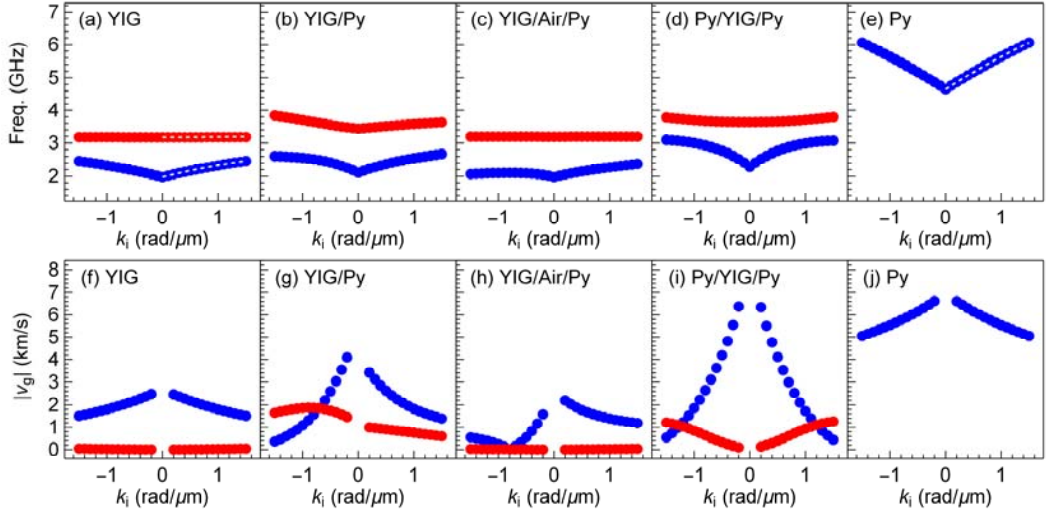

FIG. 3 Simulated dispersion relations for (a) YIG, (b) YIG/Py, (c) YIG/Air/Py, (d) Py/YIG/Py, and (e) Py at 24 mT. Corresponding group velocities for (f) YIG, (g) YIG/Py, (h) YIG/Air/Py, (i) Py/YIG/Py, and (j) Py as a function of in-plane wave vector $k_i$. The thicknesses of YIG and Py used in the simulations are 130 nm and 23 nm, respectively. The blue (red) symbols represent the MSSW (PSSW1). The dashed lines in (a) and (e) are calculated following the analytic formula in Ref. [28] for unpinned spins at boundaries.

1c is attributed to the coupling between these two modes. The anti-crossing behavior has been studied extensively in other bilayer systems and we do not further discuss this aspect [13,14]. In the following, MSSW and PSSW1 refer to the modes attributed to the YIG otherwise stated.

Next we extract $v_g$ from the two lowest lying branches in Fig. 1b and 1c. Overall, $v_g$ is found to decrease with increasing absolute value of the magnetic field as shown in Fig. 2. Specifically, at 24 mT, $\Delta f_{MSSW}$ for bare YIG and YIG/Py amounts to 65 MHz (Fig. 1e) and 117 MHz (Fig. 1g), respectively. Considering the relevant signal-to-signal line separation $d$, we obtain the corresponding $v_{MSSW}$ of 1950 m/s and 2100 m/s for the bare YIG and YIG/Py, respectively, which are similar (Fig. 2a). $\Delta f_{PSSW1}$ for the bare YIG and YIG/Py corresponds to 12 MHz and 91 MHz, which provides $v_{PSSW1}$ = 360 m/s and 1640 m/s, respectively. Here, the YIG/Py shows $v_{PSSW1}$ which is larger by a factor of 4.6 compared to that of the bare YIG. At a smaller field of 10 mT, the enhancement factor for $v_g$ amounts to even 6.5 (Fig. 2b). We note that Py is not in magnetic resonance in the frequency regime. The ferromagnetic resonance frequency of Py is about 4.6 GHz at 24 mT (Fig. 3e) and $v_g$ of Py is between 5 and 7 km/s (Fig. 3j). The signal strength of PSSW1 in YIG/Py is more than three orders of magnitude larger than that of the bare YIG (Fig. 1f and Fig. 1d, respectively). The results show that both the group velocity and excitation amplitude of PSSW1 are enhanced by the presence of the Py top layer.

The amplitudes of MSSWs are similar for YIG and YIG/Py (Fig. 1e and Fig. 1g, respectively). Based on the amplitudes, we can estimate the decay lengths for MSSWs in YIG and YIG/Py at 24 mT [26,27]. The analysis suggests that the Py layer deposited on top of YIG does not decrease the decay length.

IV. Simulation

In Fig. 3a and 3b, we show numerically calculated dispersion relations for MSSW and PSSW1 of bare YIG and YIG/Py samples at 24 mT. The solid lines in Fig. 3a reflect analytic formulas of Ref. [28] and they are consistent with the simulated frequencies. It is found that the top Py layer shifts the spin-wave frequencies of YIG to larger values and makes branches asymmetric with respect to the in-plane wave vector component $k_i$. In Fig. 3f we depict the calculated $v_{MSSW}$ (blue symbols) and $v_{PSSW1}$ (red symbols) of the bare YIG film. The $v_{PSSW1}$ is more than an order of magnitude smaller compared to $v_{MSSW}$. This is no longer true for the YIG/Py sample (Fig. 3g). Here $v_{PSSW1}$ (red symbols) has increased by roughly an order of magnitude. Also $v_{MSSW}$ is larger compared to that of the bare YIG for small wave vectors. The group velocities of YIG/Py calculated for $k_i = -0.5$ rad/μm amount to $v_{PSSW1} = 1750$ m/s and $v_{MSSW} = 2600$ m/s. These values are in good agreement with the measured values considering that the simulation was performed based on the assumed parameters and boundary conditions.

In addition, we simulate the case where the exchange coupling between YIG and Py is eliminated by replacing the 2 nm thick intermediate layer with an air layer. With the insertion of an air layer, the MSSW dispersion is still asymmetric while the dispersion of PSSW1 is flat. The corresponding velocity $v_{PSSW1}$ is similar to that of bare YIG (Fig. 3h). The result evidences that the direct contact between YIG and Py with exchange coupling is essential to observe an enhanced group velocity of PSSW1. To further investigate the enhancement mechanism we simulated a Py/YIG/Py layer. Here the exchange coupling induces forced oscillations in both Py layers, however the dynamic dipolar field from the Py layers is expected to exhibit a symmetric profile. In Fig 3i, we show the calculated group velocities. $v_{PSSW1}$ increases linearly with $k$. It approaches zero for $k \to 0$ consistent with the exchange-dominated dispersion relation ($A_{ex}k^2$). These characteristics suggest that the exchange effect is dominant for Py/YIG/Py layer. They are not observed for the asymmetrically designed bilayer YIG/Py (red dots in Fig. 3g) where $v_{PSSW1}$ is finite for $k \to 0$. In Fig. 3g the group velocity is finite for $k \to 0$ showing that the asymmetry of dynamic dipolar field is the dominant effect for an enlarged $v_{PSSW1}$ in YIG/Py.

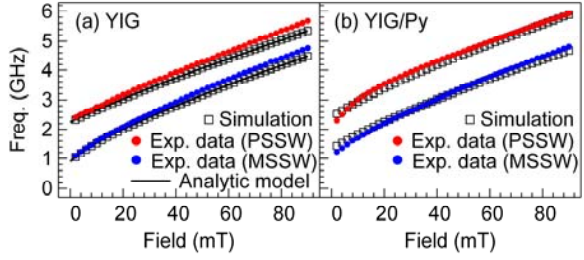

FIG. 4 Field dependence of MSSW and PSSW1 frequencies for (a) bare YIG and (b) YIG/Py bilayer. Blue and red dots are the frequencies extracted from measurement. The open black squares represent the results from the simulation. Black solid lines in (a) are the analytic formula in Ref. [28] for unpinned spins at boundaries.

Using simulated dispersion relations we calculated the group velocity at each magnetic field (Fig. 2). The solid and dashed lines in Fig. 2 represent the results for YIG and YIG/Py, respectively. The simulation reproduces the decreasing trend of group velocities with magnetic field, which is consistent with spin waves in the dipolar regime [29]. The remaining discrepancies between experiment and simulation might be attributed to the following aspects: (1) the exact pinning conditions and exchange constant, which can be influenced strongly by the quality of YIG/Py interface, are not known, and (2) possibly a complex spin configuration might exist at the interface between YIG and Py at low magnetic fields [30].

In Fig. 4a and 4b, we show the field dependence of spin wave frequencies at $k_i = -0.5$ rad/μm. The numerically calculated values (open symbols) agree very well with the analytic formula (black line) in Fig. 4a. Still, the measured frequencies (black symbols) show slight discrepancies which might be attributed to a different gyromagnetic ratio in our sample compared to the literature value. The finite conductivity of Py layer can also affect the frequency and intensity of spin wave modes in the bilayer system [31–36]. Based on the analytic method in Ref. [31], we estimated the frequency shift of spin waves in YIG in case of a 23 nm thick conductor mimicking Py. The shift amounted to less than 0.3 MHz at 24 mT (we assumed $k < 0.5$ rad/μm and a conductivity of $7 \times 10^6$ S/m). This value is roughly three orders of magnitude smaller than the linewidth of YIG resonances induced by our CPW. Following this estimation we did not take into account the conductivity of Py in our simulations.

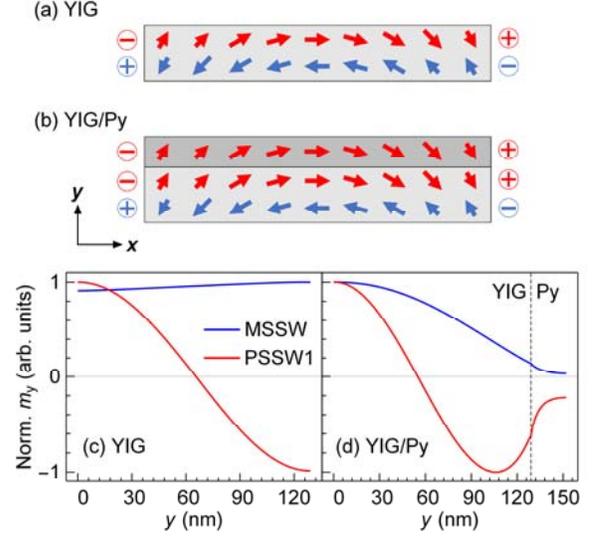

FIG. 5 Illustration of the spatial distribution of dynamic magnetization components for PSSW1 in (a) YIG and (b) YIG/Py. The induced magnetic charges are indicated by red and blue circles. The net magnetic charges cancel in (a) for YIG but not for (b) YIG/Py, creating a finite dynamic dipolar field. Simulated line profiles of $m_y(y)$ at 24 mT for (c) YIG and (d) YIG/Py. The blue and red solid lines represent MSSW and PSSW1, respectively. The black vertical dashed line in (d) indicates the boundary between YIG and Py. Maxima in $m_y(y)$ curves are normalized to one facilitating the comparison.

Now we discuss the mode profiles simulated at 24 mT (Fig. 5). We display the values of $m_y$, i.e., the dynamic magnetization component perpendicular to the film. The small amplitude variation along $y$ direction in Fig. 5c reflects the non-reciprocal character of the MSSW in a bare YIG film. While the PSSW1 in bare YIG has a perfectly anti-symmetric profile (red curve in Fig. 5c), profiles of MSSW and PSSW1 in the YIG/Py bilayer are found to display a large asymmetry along $y$ direction (Fig. 5d). Similar spin wave mode profiles were suggested in a recent work performed on YIG/CoFeB bilayer system [14]. The impact on group velocities were not discussed however. For the PSSW1 in YIG/Py, we find spin precession also in Py, not only in YIG. Though weak, the spin precession in Py is expected to alter the symmetry of dynamic dipolar field in YIG. Figure 5a and 5b illustrate the dynamic magnetization components and induced magnetic charges. The net dipolar field due to the magnetic charges in YIG cancels out because of anti-phase spin-precessional motion of top and bottom segments of YIG (Fig. 5a). The change in wave vector does not affect the cancellation effect. Therefore the PSSW1 dispersion relation is flat. The situation is

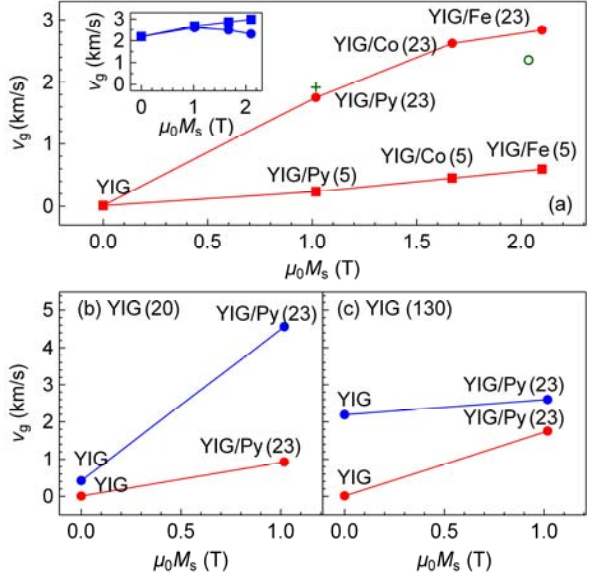

FIG. 6 (a) The simulated $v_{PSSW1}$ for different magnetic metal top layers. The disks and squares represent bilayers with YIG thickness of 130 and 20 nm, respectively. The inset in (a) shows the corresponding $v_{MSSW}$. The cross and open circle represent the YIG/Py (23) with doubled exchange constant and doubled magnetization values, respectively. The group velocities for (b) 20 nm thick YIG and (c) 130 nm thick YIG are plotted for bare YIG and YIG/Py bilayer cases. The blue (red) symbols represent $v_{MSSW}$ ($v_{PSSW1}$). The numbers in parentheses represent the layer thickness in nanometers.

different for YIG/Py. The net dynamic dipolar field from the magnetic charges does not cancel out due to the intentionally added top layer of Py (Fig. 5b). The dipolar field in PSSW1 for YIG/Py changes with wave vector similar to the MSSW in YIG [37]. Thereby $v_{PSSW1}$ in YIG/Py depends on $k$ and is enlarged.

## V. Discussion

Motivated by our experimental observations, we further explored different YIG/magnetic metal bilayer systems using our numerical simulation model. Figure 6a shows the simulation results of $v_{PSSW1}$ with different magnetic metal layers on top of 130 nm thick (filled circles) and 20 nm thick YIG (filled squares). We note that the $v_{PSSW1}$ greatly increases by the top magnetic layer while the $v_{MSSW}$ increases only slightly or decreases. The thicker the magnetic metal layer, the more the $v_{PSSW1}$ enlarges. Among the three magnetic layers, iron turns out to be the most effective material. The largest enhancement of 260 was found when adding 23 nm thick Fe to 130 nm thick YIG in Fig. 6a. Two additional simulations were made in which we artificially modified the magnetic properties of Py: one with doubled exchange constant, $A_{Py}$ (cross) and one with doubled saturation magnetization, $M_{Py}$ (open circle) in Fig. 6a. Compared to the original Py, there was an increase of a factor of 1.1 (1.3) for the case of doubled $A_{Py}$ (doubled $M_{Py}$). The result suggests that the magnetization $M_{Py}$ of the top layer is more effective to increase $v_{PSSW1}$ than the exchange constant $A_{Py}$. Simulation results obtained for different YIG thicknesses provide further insights on the origin of the observed group velocity enhancement. We increase the frequency separation between the MSSW and PSSW1 branches by reducing the YIG thickness. For 20 nm thick bare YIG, PSSW1 lies in the 40 GHz range at 24 mT while MSSW resides still in the 2 GHz range, where $v_{MSSW}$ is about 420 m/s (Fig. 6b). This value is smaller than that in the 130 nm thick YIG (Fig. 6c) as the group velocity scales nearly with the thickness [38]. However, the $v_{MSSW}$ increases significantly to about 4500 m/s when adding the top Py layer. The enhancement factor amounts to about 10 for $v_{MSSW}$ and 100 for $v_{PSSW1}$ when adding the 23 nm thick Py layer to 20 nm thick YIG (Fig. 6b). This can be compared with Fig. 6c displaying the simulation results for 130 nm thick YIG. Here, $v_{MSSW}$ increases by a factor of 1.2 while $v_{PSSW1}$ increases by a factor of 170 when adding the Py layer to the 130 nm thick YIG. The result suggests that mode repulsion can contribute to the modification of group velocity. Overall we suggest the hybridization of spin precessional motion in YIG and Py to enhance dipolar effects for magnetostatic modes and thereby enlarge their velocities. The enhanced dipolar interaction reported in this work is also advantageous to promote the recently presented concept of a directional coupler [39]. Here, dipolar effects between two separated but closely spaced YIG-based magnonic waveguides are key to provide an efficient coupling and optimum performance.

## VI. Conclusion

In summary, we investigated experimentally the propagation characteristics of MSSW and PSSW in a YIG/Py bilayer and compared it with the ones of a bare 130 nm thick reference YIG film. Supported by simulations we found that the PSSW group velocity in YIG/Py was significantly enhanced compared to that of the bare YIG. The spin wave mode profiles obtained from the numerical simulations suggested that the exchange coupling between YIG and Py caused the dynamic magnetization in Py to perform a forced precessional motion. This motion enhanced dipolar effects for the resonant PSSW modes and thereby their velocity considerably. A significant enhancement of group velocities is predicted for also MSSWs in a bilayer of 20 nm thick YIG and 23 nm thick Py. This work paves the way to optimize bilayers for directional couplers and multi-frequency magnonic devices which provide multiple spin wave modes propagating with high group velocities.

## VII. Acknowledgement

The research was funded by the EPFL COFUND project No. 665667 (EU Framework Programme for Research & Innovation (2014-2020), DU 1427/2-1 (Deutsche Forschungsgemeinschaft), Era.Net RUS Plus (TSMFA), and by SNSF via IZRPZ0_177550.